\documentclass[11pt]{article}
\usepackage[T1]{fontenc}
\usepackage{lmodern}
\usepackage[margin=1in]{geometry}
\usepackage{amsmath,amssymb,amsthm}
\usepackage{mathtools,microtype}
\usepackage{tikz}
\usetikzlibrary{arrows.meta,calc}
\usepackage{float,enumitem,placeins}
\usepackage{algorithm}
\usepackage{algpseudocode}
\usepackage[colorlinks=true,linkcolor=blue!50!black,urlcolor=blue!50!black,citecolor=blue!50!black]{hyperref}
\newtheorem{theorem}{Theorem}[section]
\newtheorem{lemma}[theorem]{Lemma}

\theoremstyle{definition}
\newtheorem{definition}[theorem]{Definition}
\theoremstyle{remark}
\newtheorem{remark}[theorem]{Remark}
\newcommand{\OPT}{\operatorname{OPT}}
\newcommand{\len}{\operatorname{len}}
\newcommand{\R}{\mathbb R}
\newcommand{\norm}[1]{\left\lVert#1\right\rVert}
\definecolor{routeblue}{RGB}{24,92,150}
\definecolor{routeorange}{RGB}{190,92,20}

\title{\bf Two-Robot Evacuation from a Unit Disk in the Segment Visibility Model}
\author{Evangelos Kranakis\\
School of Computer Science, Carleton University\\
Ottawa, Ontario, K1S 5B6, Canada}
\date{}

\begin{document}
\maketitle
\begin{abstract}
Under \emph{segment visibility}, two robots discover a target as soon as it belongs to the segment joining their positions, even if neither robot has physically reached it. We study two unit-speed robots starting at the center of a unit disk that must both reach an exit (target) placed at an unknown point of its boundary. In our model robots are permitted to leave the disk while searching. We give a three-phase tangent-and-arc algorithm whose exact worst-case evacuation time is $2\sqrt3+\pi/3\approx4.511299$. We also prove a matching lower bound for the class of descending horizontal full-slice sweeps, allowing asymmetric robot trajectories. The proof uses an exterior path-length inequality and a feasible minimum-half-width symmetrization. Determining the optimal time over all segment-visibility algorithms remains open.

\vspace{.3cm}
\noindent
{\bf Key words and phrases.} Disk, Evacuation, Robots, Search, Segment Visibility.

\smallskip
\noindent
{\bf AMS subject classification (2020).} Primary 68W05 (nonnumerical
algorithms), 68W15 (distributed algorithms); Secondary 68R01 (discrete
mathematics in relation to computer science).
\end{abstract}

\section{Introduction}\label{sec:introduction}
Searching for a hidden exit is a basic coordination task for mobile robots.
Searchers must first obtain enough information to locate the exit, yet the
cost of a successful search is not merely the discovery time: when evacuation
is required, \emph{every} robot must eventually arrive there. Consequently,
a trajectory that uncovers many candidate positions quickly may still be
poor if it leaves one robot far from the discovered exit.

Communication and sensing assumptions are crucial in evacuation problems.
In the classical face-to-face model, robots can exchange
information only when they are colocated; in the sender/receiver model the robots can communicate face-to-face but also one robot can send messages wirelessly and the other can receive messages wirelessly; in the wireless model, a robot that finds an exit can inform
the others immediately. Here we investigate a different, geometric
capability. At each instant, the line segment joining the robots is treated
as a sensor, and an exit anywhere on that segment is detected by both
robots. This makes it possible to inspect a horizontal \emph{slice} of the
disk without requiring either endpoint robot to visit every point on that
slice. The resulting algorithms combine a two-dimensional coverage problem
with the geometry of the final travel to the exit.

Our aim is to give an explicit evacuation algorithm with a rigorous
worst-case guarantee and to understand what can be proved about its
optimality. A natural first step is to examine a class of descending
horizontal sweeps whose two endpoints may move independently. The
distinction between this restricted class and completely general
trajectories is an essential part of the results.

\subsection{Preliminaries, notation, and terminology}\label{subsec:prelim}
Let $O=(0,0)$ and write
\[
 \mathbb D=\{z\in\R^2:\norm z\le1\},\qquad
 \mathbb S^1=\partial\mathbb D,
\]
for the unit disk and its boundary.
An adversary places one stationary exit $E\in\mathbb S^1$; the robots do
not know its location. Robots $R_-$ and $R_+$ begin at $O$ at time $0$.

\paragraph{Search trajectories and evacuation trajectories.}
We call $p_-(t),p_+(t)\in\R^2$ the \emph{search trajectories}:
they describe where the robots move while they are searching for the
exit. To analyze different possible exit locations using the same
functions, we view these trajectories as continuing according to the
search instructions even when discovery of a particular exit would
interrupt the search. This continuation is a mathematical convention;
it does \emph{not} mean that the robots ignore an exit they detect.
The search trajectories are $1$-Lipschitz: for each $\sigma\in\{-1,+1\}$,
\begin{equation}\label{eq:speed}
 \norm{p_\sigma(t)-p_\sigma(t')}\le |t-t'|
 \qquad\text{for all times $t,t'$ in the search trajectory's domain.}
\end{equation}
In particular, the trajectories are absolutely continuous on bounded
intervals, and $\norm{p_\sigma'(t)}\le1$ for almost every $t$ in the
sense of Lebesgue measure. Thus a robot travels along any subpath of length at most the elapsed
time. Robots may take shortcuts through the disk, travel outside it,
stop, or change direction. Acceleration is not charged.

\paragraph{Segment communication and segment detection.}
For points $a,b\in\R^2$, let
$[a,b]=\{(1-\lambda)a+\lambda b:0\le\lambda\le1\}$ be the \emph{closed}
segment, including its endpoints. The sensing convention of this paper is
that the hidden exit is simultaneously revealed to both robots at the first
instant when
\begin{equation}\label{eq:detection}
 E\in[p_-(t),p_+(t)].
\end{equation}
This condition is stronger than communication only after a robot has
physically visited the exit. In particular, ``segment communication'' here
refers to this finite \emph{robot-to-robot segment} rule, not to an
unbounded geometric ray. Once the exit is known, both robots may head
straight for it at full speed.

For a fixed exit $E$, define its detection time
\[
 \tau(E)=\inf\{t\ge0:E\in[p_-(t),p_+(t)]\}.
\]
The closed-segment rule and continuity make this a first detection time
whenever detection occurs. For an actual exit $E$, the robots follow
their search trajectories $p_\pm(t)$ until time $\tau(E)$. At that
instant they discover $E$, stop searching, and travel directly to it;
their subsequent \emph{evacuation trajectories} therefore depend on
$E$ and need not coincide with the continued search trajectories.
Because the exit belongs to the connecting segment at time $\tau(E)$,
and direct travel to a known destination is optimal, the evacuation
time for the prescribed algorithm is
\begin{equation}\label{eq:objective}
 T(E)=\tau(E)+\max\!\left\{
 \norm{p_-(\tau(E))-E},\norm{p_+(\tau(E))-E}\right\}.
\end{equation}
The \emph{worst-case evacuation time} is
$T_{\mathcal A}=\sup_{E\in\mathbb S^1}T(E)$.
Let $\OPT_{\rm segment}$ be the infimum of this quantity over all admissible
deterministic algorithms. We will also define a class $\mathcal S$
of descending horizontal-sweep algorithms, allowing asymmetric robot
trajectories, and write $\OPT_{\mathcal S}$ for its optimal time.
All time bounds are absolute times from the starting point $O$.
For a unit disk, an omniscient pair of robots that knows the exit in
advance can both reach it in exactly one unit of time. Thus if we
normalize by this offline optimum, the worst-case evacuation times
reported below are numerically equal to competitive ratios.

\paragraph{Meaning of ``almost everywhere.''}
Throughout the paper this means \emph{except for a set of one-dimensional
Lebesgue measure zero} in the time or arc-length parameter. It has no
probabilistic interpretation. This is the natural convention for paths
that may contain corners or need not be differentiable at every instant;
we return to it in the proof of Lemma~\ref{lem:exterior}.

\subsection{Related work}\label{subsec:related}
The disk-evacuation problem was introduced and studied systematically by
Czyzowicz, G\k{a}sieniec, Gorry, Kranakis, Martin, and Pajak
\cite{CGGKMP14}. Their model places robots at the center of a unit disk,
with an unknown exit on its boundary, and considers both local and wireless
communication. It set the stage for work distinguishing \emph{finding} an
exit from \emph{evacuating all robots} through it.

\paragraph{Face-to-face communication (F2F).}
In the F2F model a robot discovers the exit by physically reaching it,
and two robots may exchange information only when colocated. This
restriction can require an informed robot to intercept an uninformed one
before both can evacuate. Czyzowicz et al.\ \cite{CGKNOV20} developed
forced-meeting strategies and proved an upper bound of $5.628$ and a
lower bound of $3+\pi/6+\sqrt3\approx5.255$ for the classical two-robot
setting. Brandt et al.~\cite{Brandt2017} subsequently designed a
non-forced-meeting detour and reduced the upper bound to $5.625$.
Disser and Schmitt~\cite{DS19} obtained $5.6234$ using a second
interior detour. These are results for a different sensing and
communication model, and their lower bounds do not transfer to the
present segment model.

\paragraph{Wireless communication.}
In the wireless (WiFi) model an exit is discovered when a robot
physically reaches it, but its location can then be communicated
instantaneously to every robot. For two unit-speed robots starting
at the center of a unit disk, Czyzowicz et al.~\cite{CGGKMP14}
established the optimal WiFi evacuation time
$1+2\pi/3+\sqrt3$. Their analysis illustrates the interplay between
exploring the boundary and traversing a final chord to the exit.
In contrast, in our segment-visibility model the connecting segment
may detect a boundary exit before either endpoint robot reaches it.

\paragraph{Asymmetric sender--receiver communication.}
Georgiou, Giachoudis, and Kranakis \cite{GGK22} studied two robots with
different communication capabilities: a sender can transmit but not
receive wirelessly, while a receiver can receive but not transmit
wirelessly; both can also exchange information when they meet. Their
sender--receiver (SR) disk-evacuation algorithms include a worst-case
upper bound of $\pi+2$. This is another example of how changes in
information flow reshape optimal trajectories.

\paragraph{Other evacuation settings and surveys.}
The disk problem has generated several related models. Priority
(queen-and-servants) evacuation asks only that a distinguished robot
reach the exit; Czyzowicz et al.~\cite{Priority2018} obtain bounds for
this asymmetric objective. Czyzowicz et al.~\cite{MultipleExits2018}
study evacuation on a circle with several exits and uncertain robot
locations. Fault-tolerant disk evacuation, where one of three robots
may crash or behave in a Byzantine manner, was investigated by
Czyzowicz et al.~\cite{Faulty2017}. Wireless evacuation in other
planar domains, including equilateral triangles and squares, was
studied by Czyzowicz et al.~\cite{TriangleSquare2015}. For a broader account of group
search, continuous mobile-agent evacuation, and the relevant
communication models, see the survey by Czyzowicz, Georgiou, and
Kranakis~\cite{GroupSurvey2019}. These variants change the information,
objective, or failure assumptions, and their bounds should not be used
as bounds for our model without a separate reduction.

In the segment-visibility model, discovery is triggered by
membership in the connecting segment, rather than solely by physical
contact, and robots may search outside the unit disk. Consequently,
numerical bounds for F2F, WiFi, and SR evacuation cannot in general
be transferred to the present model without a separate argument.

\subsection{Outline of results}\label{subsec:results}
Section~\ref{sec:algorithm} presents a tangent-and-arc algorithm
whose search trajectories may leave the disk and whose \emph{exact}
worst-case evacuation time is
\begin{equation}\label{eq:mainbound}
 C_*:=2\sqrt3+\frac\pi3\approx4.5112991663.
\end{equation}
Thus $\OPT_{\rm segment}\le C_*$. 
Section~\ref{sec:lower} defines a class $\mathcal S$ of descending
horizontal full-slice sweeps allowing symmetric or asymmetric
endpoint trajectories and proves the matching result
$\OPT_{\mathcal S}=C_*$. The lower-bound proof uses an exterior
path-length inequality and a feasible minimum-half-width
symmetrization. Section~\ref{sec:conclusion} summarizes the results
and open questions. Note that we do \emph{not} claim
$\OPT_{\rm segment}=C_*$: the unrestricted lower-bound problem
remains open. 

\section{Main Algorithm}\label{sec:algorithm}

Our goal is to design two coordinated trajectories so that the
connecting segment detects every possible boundary exit, while keeping
the farther robot's distance from the exit small at discovery.
The robots may move outside the disk; this permits a full horizontal
slice to be swept by a segment whose endpoints lie beyond the circle.
The algorithm uses two straight-line phases followed by symmetric
circular arcs.

\begin{algorithm}[htbp]
\caption{Tangent-and-arc evacuation with segment detection}\label{alg:main}
\begin{algorithmic}[1]
\Require Two unit-speed robots $R_-,R_+$ at $O$; unknown $E\in\mathbb S^1$.
\State Set $A_\sigma=(\sigma/\sqrt3,1)$ and
       $B_\sigma=(\sigma\sqrt3/2,1/2)$ for $\sigma\in\{-1,+1\}$.
\Statex \textbf{Continuous interrupt (applies in every phase):}
\Statex \quad If $E\in[p_-(t),p_+(t)]$, reveal $E$ to both robots;
\Statex \quad both move along their straight segments to $E$ at speed $1$ and halt.
\State \textbf{Phase I:} Simultaneously move $R_\sigma$ straight from $O$ to $A_\sigma$ at speed $1$.
\State \textbf{Phase II:} Simultaneously move $R_\sigma$ straight from $A_\sigma$ to $B_\sigma$ at speed $1$.
\State \textbf{Phase III:} For $0\le s\le 2\pi/3$, simultaneously follow
       $p_\sigma(\sqrt3+s)=(\sigma\sin(\pi/3+s),\cos(\pi/3+s))$.
\State At $Q=(0,-1)$ the search is complete if no previous detection occurred.
\end{algorithmic}
\end{algorithm}

The parameters of Algorithm~\ref{alg:main} are as follows.
\begin{equation}\label{eq:points}
 A_\pm=\left(\pm\frac1{\sqrt3},1\right),\qquad
 B_\pm=\left(\pm\frac{\sqrt3}{2},\frac12\right),\qquad
 Q=(0,-1).
\end{equation}
The robots first move from $O$ to $A_\pm$, then along the tangent
segments $A_\pm B_\pm$, and finally along the circle from $B_\pm$ to
$Q$ on their respective sides. The connecting segment is horizontal at
every time along the search trajectories.

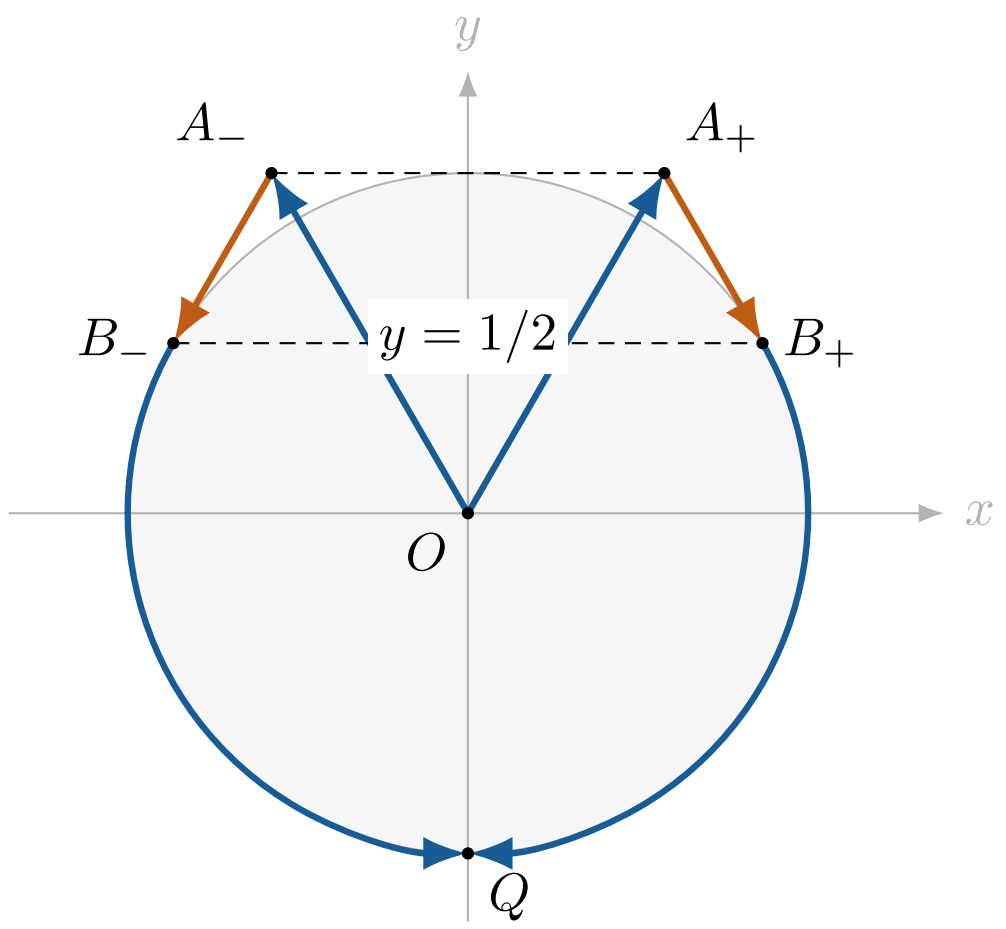
\begin{figure}[htbp]
\centering
\begin{tikzpicture}[scale=2.25,>=Latex,font=\small]
 \fill[gray!7] (0,0) circle (1);
 \draw[gray!60] (0,0) circle (1);
 \draw[->,gray!60] (-1.35,0)--(1.4,0) node[right] {$x$};
 \draw[->,gray!60] (0,-1.2)--(0,1.3) node[above] {$y$};
 \draw[very thick,routeblue,->] (0,0)--(.57735,1);
 \draw[very thick,routeblue,->] (0,0)--(-.57735,1);
 \draw[very thick,routeorange,->] (.57735,1)--(.866025,.5);
 \draw[very thick,routeorange,->] (-.57735,1)--(-.866025,.5);
 \draw[very thick,routeblue,->] (.866025,.5) arc (30:-90:1);
 \draw[very thick,routeblue,->] (-.866025,.5) arc (150:270:1);
 \draw[densely dashed] (-.57735,1)--(.57735,1);
 \draw[densely dashed] (-.866025,.5)--(.866025,.5);
 \fill (0,0) circle (.018) node[below left] {$O$};
 \fill (.57735,1) circle (.018) node[above right] {$A_+$};
 \fill (-.57735,1) circle (.018) node[above left] {$A_-$};
 \fill (.866025,.5) circle (.018) node[right] {$B_+$};
 \fill (-.866025,.5) circle (.018) node[left] {$B_-$};
 \fill (0,-1) circle (.018) node[below right] {$Q$};
 \node[fill=white,inner sep=2pt] at (0,.52) {$y=1/2$};
\end{tikzpicture}
\caption{The three search phases. The orange segments are tangents. During
the tangent phase, the horizontal connecting segment sweeps the upper cap;
the subsequent boundary arcs cover the rest of the circle.}
\label{fig:algorithm}
\end{figure}

The three search phases are illustrated in Figure~\ref{fig:algorithm}.
The phase durations are
\begin{equation}\label{eq:durations}
 |OA_\sigma|=\sqrt{\frac13+1}=\frac2{\sqrt3},\qquad
 |A_\sigma B_\sigma|=\frac1{\sqrt3},\qquad
 \len(\widehat{B_\sigma Q})=\frac{2\pi}{3}.
\end{equation}

Indeed $A_\sigma\cdot B_\sigma=1$, so
$(A_\sigma-B_\sigma)\cdot B_\sigma=0$ and the middle segment is
tangent at $B_\sigma$. The two straight phases finish at absolute times
$2/\sqrt3$ and $\sqrt3$, respectively.

\subsection{Trajectories, feasibility, and coverage}

For clarity we specify the complete search trajectories, which are followed until discovery interrupts the search.
In Phase I, at absolute time $0\le t\le2/\sqrt3$,
\begin{equation}\label{eq:phase1path}
 p_\sigma(t)=\left(\frac{\sigma t}{2},\frac{\sqrt3t}{2}\right).
\end{equation}
The derivative has norm $\sqrt{1/4+3/4}=1$; both robots are at the same
height and separated by distance $t$. In Phase II, write
$u=t-2/\sqrt3\in[0,1/\sqrt3]$. Then
\begin{equation}\label{eq:tangentcoords}
 p_\sigma(t)=(\sigma x(u),y(u)),\quad
 x(u)=\frac1{\sqrt3}+\frac u2,\quad
 y(u)=1-\frac{\sqrt3u}{2}.
\end{equation}
Here $\norm{p_\sigma'(t)}^2=1/4+3/4=1$, and direct expansion gives
\begin{equation}\label{eq:exterior}
 x(u)^2+y(u)^2
 =1+\left(u-\frac1{\sqrt3}\right)^2\ge1.
\end{equation}
The height $y(u)$ drops continuously from $1$ to $1/2$.

\begin{lemma}[Complete detection]\label{lem:coverage}
For every exit $E\in\mathbb S^1$, Algorithm~\ref{alg:main} detects $E$
by the time the robots reach $Q$. Every movement before detection is
feasible at speed one.
\end{lemma}
\begin{proof}
Feasibility in the first two phases follows from the unit norms just
computed. In Phase III, with $\theta=\pi/3+s$, the robot position is
$(\sigma\sin\theta,\cos\theta)$. Differentiating with respect to the
elapsed time $s$ gives velocity
$(\sigma\cos\theta,-\sin\theta)$ of norm one.

Consider any level $y\in[1/2,1]$. The intersection of the unit circle
with this horizontal line consists of
$(-\sqrt{1-y^2},y)$ and $(\sqrt{1-y^2},y)$ (coincident at the top).
During Phase II, the connecting segment runs from $(-x(u),y(u))$ to
$(x(u),y(u))$. Inequality~\eqref{eq:exterior} is equivalent to
$x(u)\ge\sqrt{1-y(u)^2}$, so \emph{both} circle points at that height
belong to the connecting segment. Since $y(u)$ assumes every value in
$[1/2,1]$, this phase detects every exit in the upper cap unless it was
already detected in Phase I.

Every remaining boundary point has height at most $1/2$ and can be
written as $(\sigma\sin\theta,\cos\theta)$ with
$\theta\in[\pi/3,\pi]$ and $\sigma\in\{-1,+1\}$. Robot $R_\sigma$
passes through that point during Phase III. The endpoints of the
connecting segment are included in the detection rule, so it is
necessarily detected at that moment. At $Q$ the trajectories meet,
covering the bottom point as well. This covers the full circle.
\end{proof}

\subsection{Exact evacuation-time analysis}
\begin{theorem}[Upper bound and exact cost]\label{thm:upper}
Algorithm~\ref{alg:main} has exact worst-case evacuation time
\[
 \sup_{E\in\mathbb S^1}T(E)
 =C_*=2\sqrt3+\frac\pi3\approx4.5112991663.
\]
The worst-case exits are
$P_\pm=(\pm\sqrt3/2,-1/2)$. In particular,
$\OPT_{\rm segment}\le C_*$.
\end{theorem}
\begin{proof}
By Lemma~\ref{lem:coverage}, every exit is eventually detected. If $E$
lies on the segment from positions $a$ to $b$, then its distance from
either endpoint is at most $\norm{a-b}$: this follows directly from
$E=(1-\lambda)a+\lambda b$ for some $\lambda\in[0,1]$.
Thus, at any first detection time $\tau$,
\begin{equation}\label{eq:sepbound}
 T(E)\le\tau+\norm{p_+(\tau)-p_-(\tau)}.
\end{equation}
We estimate the evacuation time according to the phase in which the
exit is first detected.

\smallskip\noindent\emph{Phase I.}
At time $t\in[0,2/\sqrt3]$, formula~\eqref{eq:phase1path} gives a
separation of $t$. Hence~\eqref{eq:sepbound} implies
\[
 T(E)\le t+t=2t\le\frac4{\sqrt3}<C_*.
\]
This conclusion remains true even if the exit is detected at an endpoint
or in the interior of the segment.

\smallskip\noindent\emph{Phase II.}
The time in this phase is at most $\sqrt3$ and the separation is
$2x(u)$. The expression for $x(u)$ in~\eqref{eq:tangentcoords} is
increasing, and at the end of the phase
$2x(1/\sqrt3)=\sqrt3$. Consequently
\[
 T(E)\le\sqrt3+\sqrt3=2\sqrt3<C_*.
\]
No optimization over possible detection heights is needed for this
coarse but sufficient upper bound.

\smallskip\noindent\emph{Phase III.}
Let $\theta\in[\pi/3,\pi]$ be the common polar angle measured
clockwise from the positive vertical axis. Both robots have positions
\begin{equation}\label{eq:phase3positions}
 p_\pm(\theta)=(\pm\sin\theta,\cos\theta),
\end{equation}
and the absolute time is
\begin{equation}\label{eq:phase3time}
 t(\theta)= 
            \frac2{\sqrt3}+\frac1{\sqrt3}+\left(\theta-\frac\pi3\right)
          =\sqrt3+\theta-\frac\pi3.
\end{equation}
For a boundary exit first detected during the interior of this phase,
strict convexity of the disk implies that the exit is an endpoint of the
connecting segment: the open segment between two distinct boundary
points lies in the open disk. Thus one robot is at the exit and its
partner is at the same height on the opposite side. The
partner must traverse the horizontal chord of length $2\sin\theta$.
The evacuation time is exactly
\begin{equation}\label{eq:phase3}
 T(E)=f(\theta)
 =\sqrt3+\theta-\frac\pi3+2\sin\theta.
\end{equation}
We maximize this continuous function on the entire compact interval.
Its derivatives are
\[
 f'(\theta)=1+2\cos\theta,\qquad
 f''(\theta)=-2\sin\theta.
\]
Since $\cos\theta$ decreases over $[\pi/3,\pi]$, the derivative
$f'$ vanishes exactly at $\theta=2\pi/3$, is positive beforehand and
negative afterwards. Thus the unique maximum of the phase-three
expression occurs at $2\pi/3$, and
\[
 f(2\pi/3)
 =\sqrt3+\left(\frac{2\pi}{3}-\frac\pi3\right)
     +2\sin\left(\frac{2\pi}{3}\right)
 =2\sqrt3+\frac\pi3=C_*.
\]
The corresponding exit points are $P_\pm=(\pm\sqrt3/2,-1/2)$.
All connecting segments in Phases I and II have positive height,
whereas $P_\pm$ have height $-1/2$. Hence neither exit could have
been detected before Phase III. At angle $2\pi/3$ one robot is at the
exit and the other is at distance exactly $\sqrt3$, so their
actual evacuation time \emph{equals} $C_*$. This proves both the
upper bound and its sharpness for the stated algorithm.
\end{proof}

\section{Lower Bound}\label{sec:lower}
The upper bound above holds for all possible exit placements, without
restricting the algorithm class. To obtain a matching \emph{lower}
bound, we now restrict the form of the search trajectories.
Within this class, we show that the tangent-and-arc algorithm cannot
be improved, even if the robots may follow arbitrary curved paths or
wait during the initial part of the search.

\subsection{The descending horizontal-sweep class}\label{sec:class}
\begin{definition}[Class $\mathcal S$]\label{def:class}
A deterministic algorithm belongs to $\mathcal S$ if, after a fixed
rotation of the coordinate system, there exist offsets $a_-,a_+\ge0$
and times $0\le t_0<t_f<\infty$ such that its \emph{search}
trajectories satisfy the following conditions:
\begin{enumerate}[label=(\roman*),leftmargin=2.2em,itemsep=4pt]
\itemsep=-\parsep
\item \textbf{Initial phase.} Both robots start at $O$ and arrive
 simultaneously at $(-a_-,1)$ and $(a_+,1)$ at time $t_0$.
 Throughout $[0,t_0]$ their vertical coordinates are nonnegative.
 The initial paths need not be straight or symmetric.
\item \textbf{Descending phase.} For $t\in[t_0,t_f]$, the positions are
 \begin{equation}\label{eq:asympaths}
 p_-(t)=(-u(t),y(t)),\qquad p_+(t)=(v(t),y(t)),
 \end{equation}
 where $u(t),v(t)\ge0$ and the common height $y(t)$ is continuous
 and nonincreasing from $y(t_0)=1$ to $y(t_f)=-1$.
 Thus the connecting segment is horizontal, but its endpoints need
 not be mirror images.
\item \textbf{Full-slice coverage.} At every time in the descending phase,
 \begin{equation}\label{eq:sweepconstraint}
 u(t),\ v(t)\ge\sqrt{1-y(t)^2}.
 \end{equation}
 Consequently the connecting segment contains both intersections of
 the unit circle with the horizontal line at height $y(t)$.
\item \textbf{Movement and evacuation.} Both robots move at speed at
 most one, and after detecting the exit each travels directly to it.
\end{enumerate}
\end{definition}
This is a \emph{single} class: symmetric sweeps are the special case
$a_-=a_+$ and $u=v$. In either case, condition~\eqref{eq:sweepconstraint}
implies $u(t)^2+y(t)^2\ge1$ and $v(t)^2+y(t)^2\ge1$ during the descent.
Thus both robot paths lie outside the open disk in that phase; this is
a restriction of the algorithm class, not a physical obstacle.
The definition constrains the search trajectories, as defined in
Subsection~\ref{subsec:prelim}, rather than the paths followed after
discovery. If the exit lies on the connecting segment, both robots
leave their search trajectories and evacuate directly toward it.

\begin{lemma}[Membership]\label{lem:membership}
Algorithm~\ref{alg:main} belongs to the class $\mathcal S$. 
\end{lemma}
\begin{proof}
Take $a_-=a_+=1/\sqrt3$ and $t_0=2/\sqrt3$.
Phase I stays above the horizontal axis and ends at $(\pm1/\sqrt3,1)$.
From that instant onward the robots have the symmetric special case
$u(t)=v(t)=x(t)$ of~\eqref{eq:asympaths}. In Phase II, $y$
decreases from $1$ to $1/2$ and $x(t)^2+y(t)^2\ge1$ by
\eqref{eq:exterior}, which gives the full-slice condition
\eqref{eq:sweepconstraint}. In Phase III this condition holds
with equality because both endpoints lie on the circle.
The height $\cos\theta$ decreases from $1/2$ to $-1$, and
both speeds equal one by Lemma~\ref{lem:coverage}. Therefore
Algorithm~\ref{alg:main} satisfies Definition~\ref{def:class}.
\end{proof}

\paragraph{Example.} 

\begin{figure}[htbp]
\centering
\begin{tikzpicture}[scale=2.05,>=Latex,font=\small]
 \fill[gray!8] (0,0) circle (1);
 \draw[gray!60] (0,0) circle (1);
 \draw[->,gray!60] (-1.45,0)--(1.48,0) node[right] {$x$};
 \draw[->,gray!60] (0,-1.2)--(0,1.28) node[above] {$y$};
 \draw[very thick,routeblue,->] (0,0)--(-1,1);
 \draw[very thick,routeblue,->] (0,0)--(1,1);
 \draw[very thick,routeorange,->] (-1,1)--(-1,-1);
 \draw[very thick,routeorange,->] (1,1)--(1,-1);
 \draw[densely dashed] (-1,.45)--(1,.45);
 \fill (0,0) circle (.017) node[below right] {$O$};
 \fill (-1,1) circle (.017) node[above left] {$(-1,1)$};
 \fill (1,1) circle (.017) node[above right] {$(1,1)$};
 \fill (-1,.45) circle (.014);
 \fill (1,.45) circle (.014);
 \node[fill=white,inner sep=1pt] at (0,.57) {detected slice};
 \node[routeblue] at (-.58,.82) {initial motion};
 \node[routeorange,rotate=90] at (-1.18,0) {vertical descent};
\end{tikzpicture}
\caption{A simple example of an algorithm in the class $\mathcal S$.
After reaching the points $(-1,1)$ and $(1,1)$, the robots descend
vertically while their connecting segment sweeps all horizontal slices of
the disk.}
\label{fig:class-example}
\end{figure}
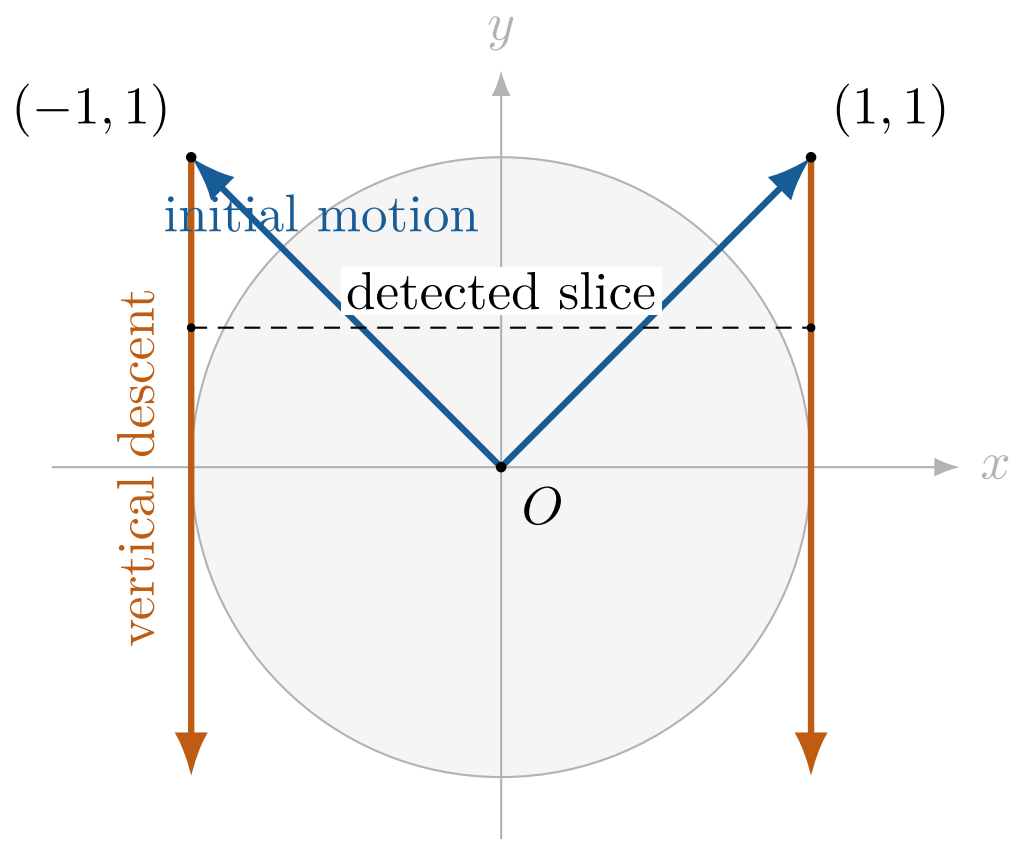

A particularly simple algorithm in $\mathcal S$ is the following
\emph{vertical sweep}. Robot $R_-$ moves along the straight segment from
$O$ to $(-1,1)$ and robot $R_+$ moves symmetrically from $O$ to $(1,1)$,
so both arrive at time $t_0=\sqrt2$. After that, they descend vertically
at unit speed along the lines $x=-1$ and $x=1$ until they reach
$(-1,-1)$ and $(1,-1)$, respectively. During the descending phase we may
write
\[
 (p_-(t),p_+(t))=((-1,y(t)),(1,y(t))),
 \qquad y(t)=1-(t-\sqrt2),
\]
for $t\in[\sqrt2,\sqrt2+2]$. Hence $u(t)=v(t)=1$ for all
such $t$, and so
\[
 u(t)=v(t)=1\ge \sqrt{1-y(t)^2},
\]
which means that every horizontal slice of the disk is fully covered. The algorithm is depicted in Figure~\ref{fig:class-example}. 

During the initial straight motion, at time $0\le t\le\sqrt2$ the
separation is $\sqrt2t$. Using~\eqref{eq:sepbound}, detection then has
evacuation time at most $(1+\sqrt2)t\le2+\sqrt2$.
Consider a boundary exit of height $y\in[-1,1]$ that was not already
detected during the initial motion. It is first detected during the
descent, at time $\sqrt2+1-y$; the farther robot is then horizontally
separated from it by $1+\sqrt{1-y^2}$. Consequently,
\begin{equation}\label{eq:vertical}
 T(E)\le\sqrt2+2-y+\sqrt{1-y^2}
 \le\sqrt2+2+\sqrt2=2+2\sqrt2.
\end{equation}
The last inequality is Cauchy--Schwarz applied to
$(-y,\sqrt{1-y^2})\cdot(1,1)$, or follows from its elementary
trigonometric maximum. Equality is attained at $y=-1/\sqrt2$, where
the exit cannot have been detected in the earlier straight motion.
Thus the \emph{exact} worst-case time of this example is
$2(1+\sqrt2)\approx4.828427$, strictly larger than $C_*$.

\subsection{An exterior path-length inequality}\label{subsec:ext}
Our lower bound will force the right-hand robot to travel from a
point above the circle to a point near its lower-right side. Instead
of assuming that the shortest exterior path is always a tangent
followed by an arc, we establish a general inequality valid for
\emph{every} rectifiable exterior path. This avoids case distinctions
when the initial offset $a$ is large.

Set
\begin{equation}\label{eq:criticalpoints}
 c=\frac{\sqrt3}{2},\qquad
 P=(c,-1/2),\qquad
 A=(a,1),\quad a\ge0,
\end{equation}
and consider the closed right exterior half-plane region
\begin{equation}\label{eq:H}
 \mathcal H=\{(x,y):x\ge0,\ x^2+y^2\ge1\}.
\end{equation}

\begin{lemma}[Exterior path-length bound]\label{lem:exterior}
If $\gamma$ is a rectifiable path from $A$ to $P$ entirely contained
in $\mathcal H$, then
\begin{equation}\label{eq:pathlower}
 \len(\gamma)\ge a+\frac{2\pi}{3}-2\arctan a.
\end{equation}
For $0\le a\le\sqrt3$, equality can be achieved by a straight
tangent segment followed by a circular boundary arc.
\end{lemma}
The preceding lemma 
concerns \emph{every} rectifiable path in the exterior
region, not only paths consisting of a tangent and a boundary arc.
Trying to identify the shortest path directly would require us to
control all possible excursions and treat different values of $a$.
Instead, we look for a scalar function $\Phi$ on $\mathcal H$ whose
spatial gradient has norm at most one. Such a function cannot increase
along any path by more than that path's Euclidean length. Indeed, for an
arc-length parameterization $\gamma:[0,L]\to\mathcal H$, where
$L=\len(\gamma)$,
\[
 \Phi(\gamma(L))-\Phi(\gamma(0))
   =\int_0^L\nabla\Phi(\gamma(s))\cdot\gamma'(s)\,ds
   \le L.
\]
The problem is thus reduced to \emph{constructing} a function whose
endpoint difference gives the desired bound. Such a function is often
called a \emph{calibration}: it is 1-Lipschitz with respect to exterior
path length (which need not equal the Euclidean distance between the
endpoints). We choose it so
that its restriction to the unit circle equals the clockwise angular
coordinate $\theta$ (hence changes by exactly the arc length), and
so that the gradient has norm one in the exterior. Writing
$\Phi(r,\theta)=\theta-F(r)$, the polar-coordinate gradient identity
\[
 \|\nabla\Phi\|^2=F'(r)^2+\frac{1}{r^2}
\]
shows that $F'(r)=\sqrt{r^2-1}/r$ and $F(1)=0$ are the natural choices.
Integrating this derivative produces the explicit $F$ in Step~1.
For the tangent-and-arc route the resulting bound is attained; for
other paths it remains a valid lower bound. The argument therefore
requires no assumption that a search trajectory is a shortest path. 

\begin{proof} (Lemma~\ref{lem:exterior})
We now construct the potential, justify the integral inequality, and
evaluate the endpoint values. Figure~\ref{fig:exterior-lemma} shows
the geometry. The path
$\gamma$ starts at the point $A=(a,1)$, remains in the exterior region
$\mathcal H$, and ends at $P=(\sqrt3/2,-1/2)$. The solid blue path is
the tangent-and-arc route that attains equality when $0\le a\le\sqrt3$.
The dashed orange curve is a generic admissible path, included only to
indicate the class of trajectories covered by the inequality. The main proof is in four steps.
\begin{figure}[htbp]
\centering
\begin{tikzpicture}[scale=2.05,>=Latex,font=\small]
 \fill[gray!8] (0,0) circle (1);
 \draw[gray!60] (0,0) circle (1);
 \draw[->,gray!60] (-.1,0)--(1.85,0) node[right] {$x$};
 \draw[->,gray!60] (0,-1.15)--(0,1.25) node[above] {$y$};
 \draw[gray!50,dashed] (0,-1.05)--(0,1.2);
 \node[gray!60] at (1.62,1.06) {$\mathcal H$};
 \coordinate (A) at (.57735,1);
 \coordinate (B) at (.866025,.5);
 \coordinate (P) at (.866025,-.5);
 \draw[very thick,routeblue] (A)--(B);
 \draw[very thick,routeblue,->] (B) arc (30:-30:1);
 \draw[very thick,routeorange,dashed,->]
    (A) .. controls (1.45,.95) and (1.58,.15) .. (1.18,-.18)
        .. controls (1.06,-.33) and (.96,-.44) .. (P);
 \draw[densely dotted] (0,0)--(B);
 \draw[densely dotted] (0,0)--(P);
 \fill (0,0) circle (.017) node[left] {$O$};
 \fill (A) circle (.017) node[above left] {$A=(a,1)$};
 \fill (B) circle (.017) node[right] {$B$};
 \fill (P) circle (.017) node[below right] {$P$};
 \draw[->] (.2,.55) arc[start angle=90,end angle=30,radius=.33];
 \node at (.5,.72) {$\theta$};
 \node at (.42,.23) {$r$};
\end{tikzpicture}
\caption{Geometry for Lemma~\ref{lem:exterior}. The path $\gamma$ must
stay in the right exterior region $\mathcal H=\{(x,y):x\ge0,
\ x^2+y^2\ge1\}$. The tangent-and-arc path $AB\cup\widehat{BP}$ is the
equality case when $0\le a\le\sqrt3$.}
\label{fig:exterior-lemma}
\end{figure}
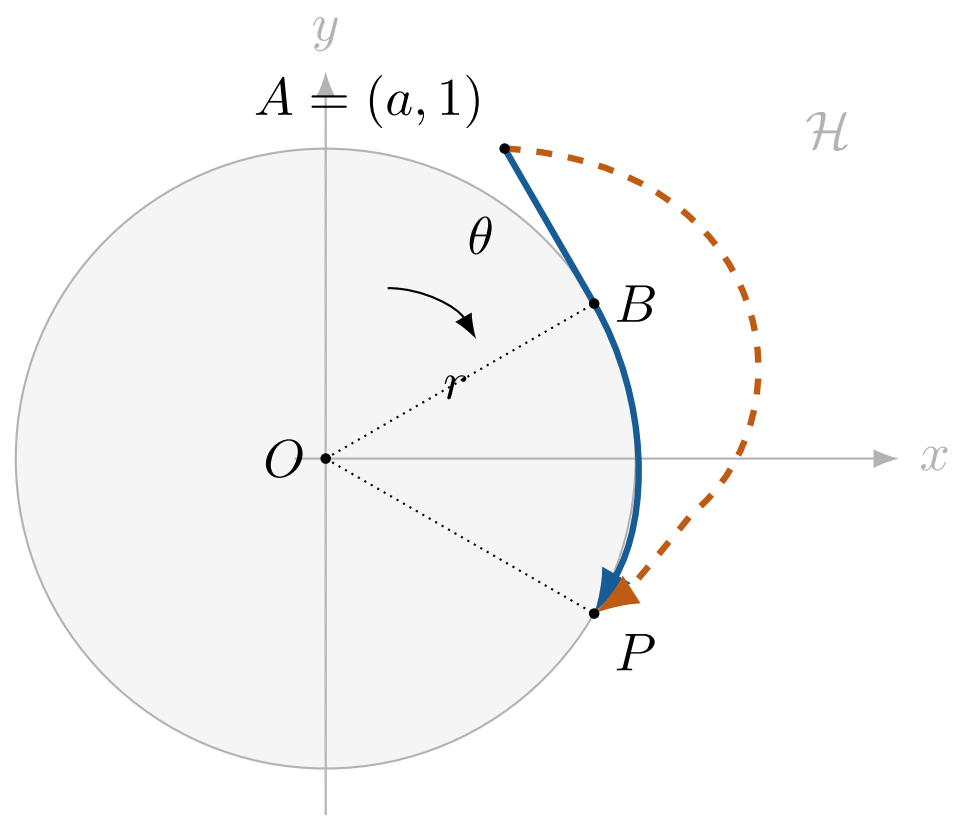

\smallskip\noindent\emph{Step 1: construct the potential.}
For a point $(x,y)\in\mathcal H$, use polar coordinates measured
clockwise from the positive vertical axis,
\[
 x=r\sin\theta,\qquad y=r\cos\theta,
 \qquad r\ge1,\quad 0\le\theta\le\pi.
\]
The right-half-plane restriction ensures that the angular coordinate
can be chosen continuously between $0$ and $\pi$. Define
\begin{equation}\label{eq:potential}
 F(r)=\sqrt{r^2-1}-\arccos\!\left(\frac1r\right),
 \qquad \Phi(r,\theta)=\theta-F(r).
\end{equation}
The two terms in $F$ are continuous at $r=1$ and $F(1)=0$.
For $r>1$, direct differentiation gives
\begin{align*}
 F'(r)
 &=\frac{r}{\sqrt{r^2-1}}
   -\frac1{r\sqrt{r^2-1}}
 =\frac{\sqrt{r^2-1}}r.
\end{align*}
In particular, $F'(r)\to0$ as $r\downarrow1$. Thus $F$ is
continuously differentiable up to the boundary $r=1$ from the exterior;
one can extend it as $F(r)=0$ for $r<1$ to obtain a continuously
differentiable function in a neighborhood of $r=1$.

In polar coordinates the squared Euclidean norm of the gradient is
\begin{equation}\label{eq:gradphi}
 \norm{\nabla\Phi}^2
 =\left(\frac{\partial\Phi}{\partial r}\right)^2
  +\frac1{r^2}\left(\frac{\partial\Phi}{\partial\theta}\right)^2
 =\frac{r^2-1}{r^2}+\frac1{r^2}=1.
\end{equation}
Note that $\norm{\nabla\Phi}=1$ at every exterior point,
including the circular boundary by continuity.

\smallskip\noindent\emph{Step 2: justify the almost-everywhere chain rule.}
Write $L=\len(\gamma)$ and parameterize $\gamma$ by arc length,
$\gamma:[0,L]\to\mathcal H$. Such a parameterization is
$1$-Lipschitz: for all $s,s'\in[0,L]$,
$\norm{\gamma(s)-\gamma(s')}\le|s-s'|$.
A Lipschitz path is absolutely continuous and has a derivative
$\gamma'(s)$ for all parameter values $s$ \emph{except a set of
one-dimensional Lebesgue measure zero}; furthermore
$\norm{\gamma'(s)}\le1$ wherever that derivative exists.
This is the one-dimensional case of Rademacher's theorem (for a complete proof, see
either Heinonen~\cite[Section~3, Theorem~3.1]{Heinonen2005} 
or Evans and Gariepy
\cite[Section~3.1.2]{EvansGariepy2015}.)

The potential $\Phi$ is continuously differentiable on the exterior
region, with local $C^1$ extensions across its boundary as explained
above. Consequently the composition $s\mapsto\Phi(\gamma(s))$ is
absolutely continuous and the ordinary chain rule applies at almost
every arc-length parameter value:
\begin{equation}\label{eq:chain}
 \frac{d}{ds}\Phi(\gamma(s))
 =\nabla\Phi(\gamma(s))\cdot\gamma'(s)
 \le\norm{\nabla\Phi(\gamma(s))}\,\norm{\gamma'(s)}
 \le1.
\end{equation}
The first inequality is the Cauchy--Schwarz inequality; the second
uses~\eqref{eq:gradphi} and the $1$-Lipschitz parameterization.
Although~\eqref{eq:chain} may fail to be \emph{defined} at isolated
corners or more general exceptional points, the exceptional set has
measure zero. The fundamental theorem for absolutely continuous
functions therefore allows us to integrate the derivative:
\begin{equation}\label{eq:integrated}
 \Phi(P)-\Phi(A)
 =\int_0^L \frac{d}{ds}\Phi(\gamma(s))\,ds
 \le\int_0^L1\,ds=L.
\end{equation}
This is why an almost-everywhere inequality is sufficient.

\smallskip\noindent\emph{Step 3: evaluate the endpoints.}
At $P=(\sqrt3/2,-1/2)$, the radial coordinate is $r=1$ and
$\theta=2\pi/3$. Therefore
\[
 \Phi(P)=\frac{2\pi}{3}-F(1)=\frac{2\pi}{3}.
\]
At $A=(a,1)$ we have
\[
 r_A=\sqrt{1+a^2},\qquad
 \theta_A=\arctan a,
\]
and, because $a\ge0$,
\begin{align*}
 F(r_A)
 &=\sqrt{(1+a^2)-1}
    -\arccos\!\left(\frac1{\sqrt{1+a^2}}\right)\\
 &=a-\arctan a.
\end{align*}
Hence $\Phi(A)=2\arctan a-a$. Substituting into
\eqref{eq:integrated} gives
\[
 L\ge\Phi(P)-\Phi(A)
 =a+\frac{2\pi}{3}-2\arctan a,
\]
which is the required inequality.

\smallskip\noindent\emph{Step 4: identify the tangency point and equality case.}
We seek a point $B$ on the right semicircle such that the segment
$AB$ is tangent to the unit circle. In the polar coordinates used
in Step~1, a circle point has the form
\[
 B=(\sin\beta,\cos\beta),\qquad 0\le\beta\le\pi.
\]
Because the circle has radius one, the tangent line at $B$ is
$\{z\in\R^2:z\cdot B=1\}$. Thus $A=(a,1)$ lies on this line if and
only if
\begin{equation}\label{eq:tangent_condition}
 A\cdot B=a\sin\beta+\cos\beta=1.
\end{equation}
Using the half-angle formulas
$\sin\beta=2\sin(\beta/2)\cos(\beta/2)$ and
$\cos\beta-1=-2\sin^2(\beta/2)$, we can rewrite
condition~\eqref{eq:tangent_condition} as
\[
 2\sin(\beta/2)\bigl(a\cos(\beta/2)
       -\sin(\beta/2)\bigr)=0.
\]
The solution $\beta=0$ corresponds to the topmost point $(0,1)$,
whose tangent line is horizontal. The second tangency point, which
lies clockwise from the top, satisfies
$\tan(\beta/2)=a$; hence
\begin{equation}\label{eq:beta_formula}
\beta=2\arctan a,\qquad
 B=(\sin\beta,\cos\beta)
   =\left(\frac{2a}{1+a^2},\frac{1-a^2}{1+a^2}\right).
\end{equation}
For $a=0$ the two tangency points coincide at $(0,1)$.

To verify directly that this point is tangent, the double-angle
identities yield
\[
 \sin(2\arctan a)=\frac{2a}{1+a^2},\qquad
 \cos(2\arctan a)=\frac{1-a^2}{1+a^2}.
\]
Substituting these expressions, one term at a time, gives
\begin{align*}
 A\cdot B
 &=a\sin\beta+\cos\beta\\
 &=a\left(\frac{2a}{1+a^2}\right)
     +\frac{1-a^2}{1+a^2}\\
 &=\frac{2a^2+1-a^2}{1+a^2}
  =\frac{1+a^2}{1+a^2}=1.
\end{align*}
Consequently $(A-B)\cdot B=A\cdot B-\norm B^2=0$,
so $AB$ is tangent at $B$. Its length is
\[
 |AB|^2=|A|^2+|B|^2-2A\cdot B
       =(1+a^2)+1-2=a^2,
 \quad\text{hence } |AB|=a.
\]

When $0\le a\le\sqrt3$, equation~\eqref{eq:beta_formula} gives
$0\le\beta\le 2\pi/3$. The clockwise circle arc from $B$ to
$P$, whose angle is $2\pi/3$, therefore has nonnegative length
$2\pi/3-\beta$ and remains in $\mathcal H$. The tangent segment
followed by this arc is an admissible path of total length
\[
 |AB|+\len(\widehat{BP})
     =a+\frac{2\pi}{3}-2\arctan a,
\]
exactly the lower bound proved in Step~3. In particular, for
$a=0$ the tangent segment has zero length and the path is the circle
arc from the top to $P$.

When $a>\sqrt3$, $\beta=2\arctan a>2\pi/3$: the calculated tangency
point $B$ lies \emph{past} $P$ in the clockwise direction. Hence the
proposed clockwise arc from $B$ to $P$ no longer has length
$2\pi/3-\beta$; this quantity would be negative. The claimed
equality construction is therefore available only for
$0\le a\le\sqrt3$. In fact, for $a\ge\sqrt3$, the straight segment
$AP$ already lies in $\mathcal H$ and is a shortest admissible
path. To see this, write its points as $P+s(A-P)$ for
$0\le s\le1$. The horizontal coordinate is nonnegative and
\[
 \norm{P+s(A-P)}^2
 =1+2s\,P\cdot(A-P)+s^2\norm{A-P}^2
 \ge1,
\]
because $P\cdot(A-P)=\tfrac{\sqrt3}{2}a-\tfrac32\ge0$.
Thus it stays outside the disk; every path from $A$ to $P$ has
length at least the Euclidean distance $|AP|$.
The inequality~\eqref{eq:pathlower} remains valid for all $a\ge0$,
but the tangent-and-clockwise-arc equality construction no longer
applies when $a>\sqrt3$. This completes the proof of Lemma~\ref{lem:exterior}.
\end{proof}

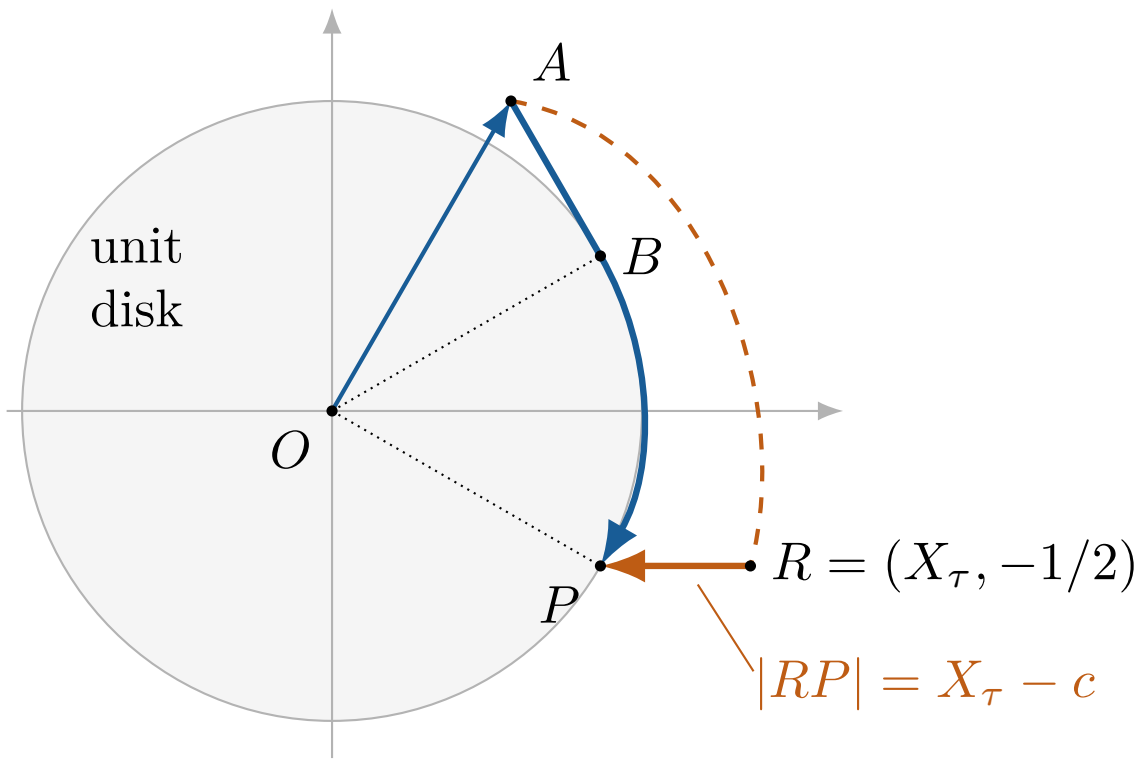
\begin{figure}[htbp]
\centering
\begin{tikzpicture}[scale=2.05,>=Latex,font=\small]
 \fill[gray!8] (0,0) circle (1);
 \draw[gray!60] (0,0) circle (1);
 \draw[->,gray!60] (-1.05,0)--(1.65,0);
 \draw[->,gray!60] (0,-1.12)--(0,1.3);
 \draw[thick,routeblue,->] (0,0)--(.57735,1);
 \draw[very thick,routeblue] (.57735,1)--(.866025,.5);
 \draw[very thick,routeblue,->] (.866025,.5) arc (30:-30:1);
 \draw[dashed,routeorange,thick] (.57735,1)
    .. controls (1.2,.9) and (1.5,.05) .. (1.35,-.5);
 \draw[very thick,routeorange,->] (1.35,-.5)--(.866025,-.5);
 \draw[densely dotted] (0,0)--(.866025,-.5);
 \draw[densely dotted] (0,0)--(.866025,.5);
 \fill (0,0) circle (.018) node[below left] {$O$};
 \fill (.57735,1) circle (.018) node[above right] {$A$};
 \fill (.866025,.5) circle (.018) node[right] {$B$};
 \fill (.866025,-.5) circle (.018) node[below left] {$P$};
 \fill (1.35,-.5) circle (.018) node[right] {$R=(X_\tau,-1/2)$};
 \draw[routeorange,thin] (1.18,-.56)--(1.36,-.84);
 \node[routeorange,anchor=west,inner sep=1pt] at (1.34,-.88) {$|RP|=X_\tau-c$};
 \node[align=center] at (-.63,.43) {unit\\disk};
\end{tikzpicture}
\caption{The augmentation used in the proof of
Theorem~\ref{thm:restricted}. The dashed orange curve is a schematic
exterior path from $A$ to $R=(X_\tau,-1/2)$; appending the orange
segment $RP$, of length $X_\tau-c$, makes a path ending at the fixed
point $P$.
The solid blue tangent-and-arc path shows equality in
Lemma~\ref{lem:exterior} when $a=1/\sqrt3$.}
\label{fig:lower}
\end{figure}

\subsection{Adversarial exit and matching lower bound}

Finally we prove the main lower bound for algorithms in the class $\mathcal{S}$.
\begin{theorem}[Optimality within the descending-sweep class]
\label{thm:restricted}
Every algorithm $\mathcal A\in\mathcal S$, 
satisfies
\[
 \sup_{E\in\mathbb S^1}T_{\mathcal A}(E)
 \ge2\sqrt3+\frac\pi3.
\]
Consequently,
\begin{equation}\label{eq:optclass}
\OPT_{\mathcal S}=2\sqrt3+\frac\pi3.
\end{equation}
\end{theorem}
\begin{proof}
Fix an arbitrary $\mathcal A\in\mathcal S$ and use the coordinates
of Definition~\ref{def:class}. Its initial offsets are $a_-,a_+$;
its descending endpoints are $(-u(t),y(t))$ and $(v(t),y(t))$.
The proof first replaces these possibly asymmetric paths by a
feasible symmetric comparison sweep, and then applies
Lemma~\ref{lem:exterior} to that sweep. The comparison is needed
only at one carefully chosen exit, not for every possible exit.

\smallskip\noindent\emph{Step 1: choose an adversarial exit and its
first detection time.}
Set
\[
 c=\frac{\sqrt3}{2},\qquad P=(c,-1/2).
\]
During the initial phase, the connecting segment lies in the
upper closed half-plane, so it cannot detect $P$. During the
descending phase every connecting segment is horizontal at height
$y(t)$, which decreases continuously from $1$ to $-1$.
Consequently
\begin{equation}\label{eq:criticaltau}
 \tau=\min\{t\in[t_0,t_f]:y(t)=-1/2\}
\end{equation}
exists and is the first time the segment can contain $P$.
At time $\tau$, condition~\eqref{eq:sweepconstraint} gives
$u(\tau),v(\tau)\ge c$, so the segment actually contains $P$.
Thus $\tau$ is the first detection time of $P$.

\smallskip\noindent\emph{Step 2: symmetrize without increasing the
speed.}
For $t\in[t_0,t_f]$, define
\begin{equation}\label{eq:minhalfwidth}
 X(t)=\min\{u(t),v(t)\},\qquad a=\min\{a_-,a_+\}.
\end{equation}
Continuity at the transition time gives
$u(t_0)=a_-$, $v(t_0)=a_+$, and hence $X(t_0)=a$.
By full-slice coverage,
\[
 X(t)\ge\sqrt{1-y(t)^2},
\]
so the symmetric pair $(-X(t),y(t))$ and $(X(t),y(t))$ still
covers every horizontal disk slice.

The functions $u,v,y$ are absolutely continuous. Since
$X=(u+v-|u-v|)/2$, $X$ is absolutely continuous too.
At almost every time with $u<v$ we have $X'=u'$, and at almost
every time with $v<u$ we have $X'=v'$. On the set $u=v$ the
derivatives $u'$ and $v'$ agree almost everywhere (at density
points of their equality set), so $X'$ again equals a permitted
horizontal velocity. Therefore, using the unit-speed conditions
for the original two robots,
\begin{equation}\label{eq:minspeed}
 X'(t)^2+y'(t)^2\le1\quad\text{for almost every }t.
\end{equation}
This proves feasibility of the symmetric descending trajectories,
including at switch points of the minimum.

Moreover, the original robots reach $(-a_-,1)$ and $(a_+,1)$
by time $t_0$, so
\[
 t_0\ge\max\{\sqrt{1+a_-^2},\sqrt{1+a_+^2}\}
 \ge\sqrt{1+a^2}.
\]
We may therefore send two comparison robots straight from $O$
to $(-a,1)$ and $(a,1)$ over the same time interval $[0,t_0]$,
at a constant speed at most one. Their vertical coordinates
remain nonnegative. Joining these initial segments to
$(-X(t),y(t))$ and $(X(t),y(t))$ defines a feasible symmetric
comparison sweep $\mathcal A_{\min}\in\mathcal S$.

\smallskip\noindent\emph{Step 3: compare evacuation times at $P$.}
The comparison sweep and the original sweep share the same
height function $y(t)$, so they both first detect $P$ at the time
$\tau$ from~\eqref{eq:criticaltau}. Put
$u_\tau=u(\tau)$, $v_\tau=v(\tau)$ and $X_\tau=X(\tau)$.
The original evacuation time is
\[
 T_{\mathcal A}(P)=\tau+\max\{u_\tau+c,\,v_\tau-c\}
 \ge\tau+u_\tau+c\ge\tau+X_\tau+c.
\]
For the symmetric comparison sweep, the left robot is farther
from $P$ than the right robot, giving
\begin{equation}\label{eq:TP}
 T_{\mathcal A_{\min}}(P)=\tau+X_\tau+c.
\end{equation}
Hence $T_{\mathcal A}(P)\ge T_{\mathcal A_{\min}}(P)$.
We stress that this is a comparison at the \emph{single exit} $P$;
we do not assert such an inequality for all exit locations.

\smallskip\noindent\emph{Step 4: bound the comparison path to
first detection.}
At time $t_0$ the right comparison robot is at $A=(a,1)$.
The initial travel distance is $\sqrt{1+a^2}$, so
\begin{equation}\label{eq:t0bound}
 t_0\ge\sqrt{1+a^2}.
\end{equation}
Let $\gamma$ be its descending path from $A$ to
$R=(X_\tau,-1/2)$, and let $\ell=\len(\gamma)$.
Unit speed implies
\begin{equation}\label{eq:taubound}
 \tau\ge t_0+\ell\ge\sqrt{1+a^2}+\ell.
\end{equation}
The full-slice condition ensures
$X(t)^2+y(t)^2\ge1$ and $X(t)\ge0$ throughout the descent;
therefore $\gamma\subseteq\mathcal H$, the exterior
region of~\eqref{eq:H}.

\smallskip\noindent\emph{Step 5: append a horizontal segment
and apply Lemma~\ref{lem:exterior}.}
Figure~\ref{fig:lower} illustrates this construction. The path
$\gamma$ ends at $R$, while Lemma~\ref{lem:exterior} concerns
paths ending at $P$. Because $X_\tau\ge c$, append the horizontal
segment $RP$ of length $X_\tau-c$. Every point of this segment
has coordinates $(z,-1/2)$ with $z\ge c$, and hence
\[
 z^2+\tfrac14\ge c^2+\tfrac14=1.
\]
Thus $RP\subseteq\mathcal H$ and the augmented path from
$A$ to $P$ is an admissible exterior path of length
$\ell+(X_\tau-c)$. Lemma~\ref{lem:exterior} gives
\begin{equation}\label{eq:augmented}
 \ell+(X_\tau-c)\ge a+\frac{2\pi}{3}-2\arctan a.
\end{equation}
This augmentation is useful because it removes the unknown
endpoint $R$ without requiring the original trajectory to be a
shortest path.

\smallskip\noindent\emph{Step 6: combine and optimize.}
Equations~\eqref{eq:TP}, \eqref{eq:taubound}, and
\eqref{eq:augmented} yield
\begin{align*}
 T_{\mathcal A}(P)
 &\ge T_{\mathcal A_{\min}}(P)
  =\tau+X_\tau+c\\
 &\ge\sqrt{1+a^2}+\ell+X_\tau+c\\
 &\ge\sqrt{1+a^2}+a+\frac{2\pi}{3}
       -2\arctan a-(X_\tau-c)+X_\tau+c\\
 &=\sqrt3+G(a),
\end{align*}
where $2c=\sqrt3$ and
\begin{equation}\label{eq:G}
 G(a)=\sqrt{1+a^2}+a+\frac{2\pi}{3}-2\arctan a,
 \qquad a\ge0.
\end{equation}
Notice that the unknown slice half-width $X_\tau$ has canceled.
To minimize $G$, differentiate:
\[
 G'(a)=\frac{a}{\sqrt{1+a^2}}+1-\frac2{1+a^2}.
\]
Set $s=a/\sqrt{1+a^2}$, so $0\le s<1$ and
$1/(1+a^2)=1-s^2$. We obtain
\[
 G'(a)=s+1-2(1-s^2)=(2s-1)(s+1).
\]
Since $s$ increases strictly with $a$, $G$ decreases until
$s=1/2$ and increases thereafter. The unique minimizer is
$a=1/\sqrt3$. Substitution gives
\begin{align*}
 G(1/\sqrt3)
 &=\frac2{\sqrt3}+\frac1{\sqrt3}
      +\frac{2\pi}{3}-2\arctan(1/\sqrt3)\\
 &=\sqrt3+\frac\pi3.
\end{align*}
Consequently
\[
 T_{\mathcal A}(P)\ge\sqrt3+G(a)
 \ge2\sqrt3+\frac\pi3=C_*.
\]
As $P$ is a valid exit, every algorithm in $\mathcal S$ has
worst-case time at least $C_*$. Conversely, Algorithm~\ref{alg:main}
belongs to $\mathcal S$ by Lemma~\ref{lem:membership}, and its
worst-case time equals $C_*$ by Theorem~\ref{thm:upper}.
Thus $\OPT_{\mathcal S}=C_*$.
\end{proof}

\begin{remark}[Geometry of the optimal parameters]
For the minimizing offset $a=1/\sqrt3$, the equality construction
in Lemma~\ref{lem:exterior} uses tangency angle
$\beta=2\arctan(1/\sqrt3)=\pi/3$, hence tangency point
$(\sqrt3/2,1/2)$ on the right. From there the clockwise boundary
arc to the critical exit $P=(\sqrt3/2,-1/2)$ has length $\pi/3$.
The tangent-and-arc algorithm therefore realizes equality in the
geometric path inequality at the critical exit. This observation
explains its parameters, but does not establish uniqueness of the
optimal trajectory within $\mathcal S$.
\end{remark}

\subsection{Scope of the lower-bound argument}\label{subsec:scope}

Theorem~\ref{thm:restricted} proves an exact optimum for the
class $\mathcal S$ of Definition~\ref{def:class} in which
both robot endpoints may
follow different horizontal motions, and the initial offsets may
differ. The proof uses three restrictions: the initial phase stays
above the horizontal axis; the descending detecting segment is
horizontal and has a common nonincreasing height; and \emph{both}
endpoints satisfy full-slice coverage. These ensure that the
minimum-half-width construction still follows a feasible
exterior path and first detects the critical exit at the same time.

For unrestricted segment-visibility algorithms, however, the
connecting segment may rotate, the two robots may have different
heights, and paths may revisit heights or cross the open disk.
Our symmetrization need not work in such cases, and the same exit
might be detected earlier. We have not proved that symmetrization
reduces evacuation time at every exit. Hence the distinct conclusions remain
\[
 \OPT_{\rm segment}\le C_*,\qquad \OPT_{\mathcal S}=C_*.
\]
In particular, this analysis does not establish
$\OPT_{\rm segment}\ge C_*$.

\section{Conclusion}\label{sec:conclusion}
We investigated evacuation by two unit-speed robots starting at the
center of a unit disk, with an unknown exit on its boundary. The exit
is discovered when it belongs to the segment joining the two robots.
This sensing rule permits a horizontal slice to be inspected without
requiring either robot to reach both of its boundary endpoints.

Our tangent-and-arc algorithm has exact worst-case evacuation time
$2\sqrt3+\pi/3$.
A matching lower bound holds within the class $\mathcal S$ of
descending horizontal full-slice sweeps, including asymmetric endpoint
motions. The proof constructs a feasible symmetric comparison sweep
by taking the minimum of the two horizontal half-widths and then
applies an exterior path-length potential with unit-norm gradient.
The argument remains valid for paths with corners because the chain
rule is applied almost everywhere in the measure-theoretic sense.

The central open question is whether more general segment
trajectories, such as rotating or nonhorizontal sweeps, can improve
the upper bound, or whether an adversarial argument can establish
$\OPT_{\rm segment}\ge 2\sqrt3+\pi/3$ without the horizontal
full-slice restrictions. Further directions include bounded exterior
excursions, nonmonotone search heights, and weaker sensing
conventions requiring physical discovery. 

The results of this paper concern exits on the boundary $\mathbb S^1$.
As an observation about the algorithm, its search trajectories also scan
every point of $\mathbb D$: the horizontal segments in Phase II cover
all disk slices with heights in $[1/2,1]$, and Phase III covers all
slices with heights in $[-1,1/2]$. The same phase-by-phase estimates
therefore give an exact worst-case \emph{evacuation time} of $C_*$ even
if the exit may lie in $\mathbb D$, since the boundary exits already
attain $C_*$. This does \emph{not} yield a bounded competitive ratio
for that variant. Indeed, for $E_\varepsilon=(0,-\varepsilon)$ with
$0<\varepsilon<1$, the evacuation time under Algorithm~\ref{alg:main}
is
\[
 \sqrt3+\arccos(-\varepsilon)-\frac\pi3
   +\sqrt{1-\varepsilon^2},
\]
whereas the offline optimum is $\varepsilon$. The ratio diverges as
$\varepsilon\downarrow0$. Computing the competitive ratio for this interior-exit variant
remains an open problem. 

\paragraph{Acknowledgments.}
Research of Evangelos Kranakis was supported in part
by NSERC Discovery grant. The revision used AI assistance for proof
development, adversarial review, and manuscript preparation.

\end{document}